\documentclass{elsart}

\usepackage{latexsym,psfrag,graphicx}
\usepackage{amssymb}

\usepackage[numbers]{natbib}

\begin{document}

\begin{frontmatter}

\title{Two-dimensional flows of foam: drag exerted on circular obstacles and dissipation}
\author[label1]{Benjamin Dollet}
\author[label2]{Florence Elias}
\author[label1]{Catherine Quilliet}
\author[label1]{Arnaud Huillier}
\author[label3]{Miguel Aubouy}
\author[label1]{Fran\c cois Graner\corauthref{cor1}}
    \address[label1]{Spectrom\'etrie Physique\thanksref{label4}, 140 rue
de la Physique, BP 87, 38402 Saint Martin d'H\`eres Cedex, France}
    \address[label2]{Laboratoire des Milieux D\'esordonn\'es et
H\'et\'erog\`enes\thanksref{label5}, case 78, 4 place Jussieu,
75252 Paris Cedex 05, France}
    \address[label3]{SI3M, DRFMC, CEA, 38054 Grenoble Cedex 9,
    France}
\corauth[cor1]{Author for correspondence, fax +33 4 76 63 54 95,
\texttt{graner@ujf-grenoble.fr}.}
    \thanks[label4]{UMR 5588 CNRS and Universit\'e Joseph Fourier.}
    \thanks[label5]{F\'ed\'eration de Recherche FR 2438
"Mati\`ere et Syst\`emes Complexes", Paris, France.}

\begin{abstract}
A Stokes experiment for foams is proposed. It consists in a two-dimensional
flow of a foam, confined between a water subphase and a top plate, around a
fixed circular obstacle. We present systematic measurements of the drag exerted
by the flowing foam on the obstacle, \emph{versus} various separately
controlled parameters: flow rate, bubble volume, solution viscosity, obstacle
size and boundary conditions. We separate the drag into two contributions, an
elastic one (yield drag) at vanishing flow rate, and a fluid one (viscous
coefficient) increasing with flow rate. We quantify the influence of each
control parameter on the drag. The results exhibit in particular a power-law
dependence of the drag as a function of the solution viscosity and the flow
rate with two different exponents. Moreover, we show that the drag decreases
with bubble size, increases with obstacle size, and that the effect of boundary
conditions is small. Measurements of the streamwise pressure gradient,
associated to the dissipation along the flow of foam, are also presented: they
show no dependence on the presence of an obstacle, and pressure gradient
depends on flow rate, bubble volume and solution viscosity with three
independent power laws.
\end{abstract}

\end{frontmatter}

\section{Introduction}

Liquid foams, like colloids, emulsions, polymer or surfactant solutions, are
characterised by a complex mechanical behaviour. Those systems, known as soft
complex systems, are multiphasic materials. Their constitutive entities are in
interaction, generating internal structures, which cause the diversity in the
fluid rheological behaviour \cite{Larson1999}. Liquid foams are convenient
model experimental system for studying the interplay between structure and
rheology, since their internal structure can be easily visualised and
manipulated.

Liquid foams are made of polyhedral gas bubbles separated by thin liquid
boundaries forming a connected network. The liquid phase occupies a small
fraction of the volume of the foam (several percent). The mechanics of liquid
foams is rich: foams are elastic, plastic or viscous depending on the applied
strain and strain rate \cite{Weaire1999}. This behaviour has been shown in
rheological experiments performed on three-dimensional (3D) foams
\cite{Mason1995,Mason1996,Cohen-Addad1998,Saint-Jalmes1999}; models have been
built to account for this diversity of rheological behaviour
\cite{Reinelt2000,Sollich1997,Durian1995,Cates2004}. However, the visualisation
of the foam structure is technically difficult in 3D
\cite{Monnereau1999,Prause1995}, although progress have been made recently
\cite{Lambert2004}. Moreover, the drainage of the liquid phase due to gravity
may occur in 3D, making the fluid fraction and therefore the rheological moduli
of the foam inhomogeneous \cite{Koehler2001}. An inhomogeneous liquid volume
fraction of the foam may also cause an inhomogeneous coarsening of the foam,
thus an inhomogeneous repartition of the bubble size.

For all these reasons, the mechanics of foams has been studied in two
dimensions, where the direct visualisation of the structure is easier, and no
gravity-driven drainage occurs if the system is horizontal. The system is then
either a true 2D system like a Langmuir foam \cite{Losche1983,Courty2003}, or
quasi 2D system constituted by a monolayer of bubbles, either at the free
surface of the solution (bubble raft \cite{Abdelkader1998,Pratt2003}), or
confined between two horizontal transparent plates
\cite{Debregeas2001,Asipauskas2003}, or between the surface of the solution and
an upper horizontal transparent plate \cite{Smith1952,Vaz1997}. The deformation
and motion of individual cells have been forced and studied in different flow
geometries: simple shear \cite{Abdelkader1998}, flow in a constriction or
around an obstacle \cite{Asipauskas2003}, Couette flow
\cite{Debregeas2001,Pratt2003}. Some authors have been particularly interested
in the dynamics of bubble rearrangements during the flow: the spatial
distribution of the rearrangements \cite{Abdelkader1998,Debregeas2001}, the
stress relaxation associated with the rearrangements \cite{Pratt2003}, the
deformation profile \cite{Janiaud}, the averaged velocity
\cite{Debregeas2001,Asipauskas2003}. However, no mechanical measurement has
been performed in those last studies.

In this paper, we study the mechanics of a foam flowing in relative
displacement with respect to an obstacle, at a constant velocity. In a
Newtonian liquid at low Reynolds number, the force would vary linearly with the
foam-obstacle relative velocity, the proportionality factor being linked to the
liquid viscosity and the size of the obstacle. This experiment gives then
information on the "effective viscosity" of a flowing foam. Such a Stokes
experiment has first been performed in a 3D coarsening foam by Cox et al.
\cite{Cox2000}. Here, the force exerted by the quasi 2D foam on the obstacle is
measured, as a function of the flow velocity, in a 2D geometry. A similar
experiment has been performed recently to investigate the elastic regime of a
2D foam and measure the foam shear modulus \cite{Courty2003}. In the
experiments presented here, the foam flows permanently around the obstacle, and
the stationary regime is investigated. The system used is a monolayer of soap
bubbles confined between the surface of the solution and a horizontal plate.
This allows measuring accurately forces exerted on the obstacle and dissipation
along the flow, and varying easily the foam internal parameters such as the
viscosity of the solution, the bubble size, and the geometry of the obstacle.

The article is organised as follows. The experimental material and methods are
presented in section II, the results are shown in section III for drag, and in
section IV for dissipation. These results are discussed in section IV, and
conclusions are exposed in section V.

\section{Materials and methods}

\subsection{Foam production}

\begin{figure}
\begin{center}
\includegraphics[width=8cm]{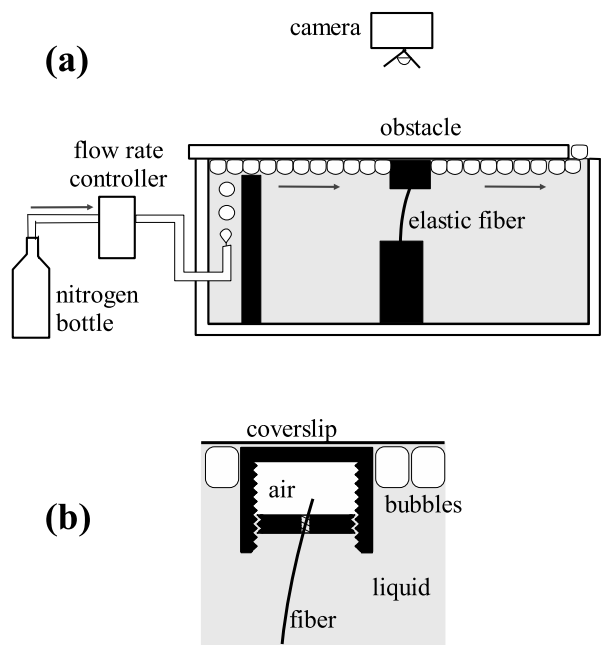}
\caption{\label{Setup+flotteur} (a) Experimental setup. The arrows
indicate the flow of gas and foam. (b) Detailed sketch of the
obstacle.}\end{center}
\end{figure}

\begin{figure}
\begin{center}
\includegraphics[width=12cm]{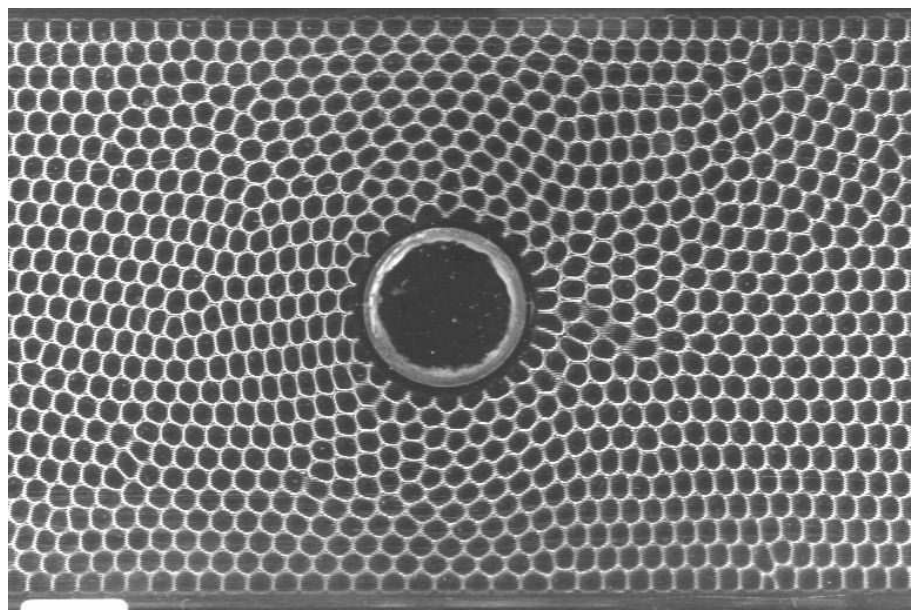}
\caption{\label{Image} Photo of foam flowing from left to right around a
circular obstacle of diameter 30 mm. The bubble size is 16.0 mm$^2$ (note the
monodispersity of the foam), and the flow rate is 174 mL/min. The walls of the
channel (width 10 cm) are visible at the top and bottom of the picture. The
stretching and shearing of bubbles due to the presence of the obstacle is
clearly visible around the obstacle. The surface of the observed field is
$15.4\times 10.2$ cm$^2$, and 1 pixel side equals 0.20 mm.}\end{center}
\end{figure}

The experimental setup is presented on Fig. \ref{Setup+flotteur}(a). The
experiments are performed in a glass channel of 110 cm length, 10 cm width and
10 cm depth. The soap solution is a solution of commercial dish-washing fluid
(1\% in volume) in purified water, with added glycerol when the viscosity needs
to be varied (subsection \ref{SectionViscosity}). The surface tension of the
solution is $\gamma=31$ mN/m. At the beginning of each experiment, the channel
is filled with the solution, with a 3.50 mm gap between the liquid surface and
the coverslip. The foam is produced by blowing bubbles of nitrogen in the
solution, at one end of the channel, in a chamber bounded by a barrier which
allows a single monolayer of bubbles to form. The continuous gas flow makes the
foam flow along the channel, between the surface of the solution and the
coverslip. A typical image of the flowing foam observed from above is displayed
in Fig. \ref{Image}.

\subsection{Obstacle and force measurements}

The obstacle stands in the middle of the channel. It is a buoyant mobile
plastic object connected to a fixed base by a soft glass fiber. The bottom
extremity of the fiber is rigidly fixed. Its top extremity simply passes
through a hole drilled in the bottom of the obstacle (Fig
\ref{Setup+flotteur}(b)). Therefore, the fiber can slide inside the
horizontally moving obstacle, without applying any undesirable vertical force.
Moreover, the fiber is lubricated by the liquid, which avoids solid friction
against the obstacle.

The horizontal force $F$ exerted by the foam on the obstacle tends to pull it
streamwise; it is balanced by the horizontal drawback force $F_d$ from the
elastic fiber, whose deflection is designed by $X$. The calculation of this
force is classical in the theory of elasticity; for a given displacement $X$
from the position at rest of the mobile part, the drawback force writes
\cite{Landau1986}: $F_d=-KX$, where the rigidity $K$ writes: $K = 3\pi
ED^4/64L^3$, with $D=240$ $\mu$m the fiber diameter, $L$ its vertical length
and $E$ its shear modulus. The fiber has been calibrated by measuring its
deflection under its own weight, giving the value of the shear modulus : $E =
6.67\times 10^{10}$ Pa. This value is compatible with typical values of the
shear modulus of glass: $6$--$7\times 10^{10}$ Pa. We use two different fibers,
of vertical length $L=34.8$ mm and $L=42.4$ mm, depending on the magnitude of
the force to measure. We have checked that for given experimental conditions,
the same force is measured with both fibers (data not shown). The displacement
is measured by tracking the position of the obstacle with a CCD camera placed
above the channel: the actual position of the obstacle is given by the
coordinates of its center, obtained by image analysis. The position of the
center of the obstacle is known with a precision of 0.02 mm, much lower than
the typical displacement (5 mm to 1 cm). When the obstacle has reached a
stationary position under flow, the drawback force exactly compensates the
force exerted by the foam, which is then directly deduced from the measured
displacement.

The obstacle is in contact with the coverslip. This is necessary for the foam
to flow around the obstacle and not above, but this may induce friction.
Nevertheless, in the setup presented here, the obstacle is in contact with a
single plate; this reduces the friction in comparison with an experiment
performed in a Hele-Shaw cell, where the foam is confined between two plates.
Furthermore, the obstacle is constituted by a hollow part closed by a
watertight screw (Fig. \ref{Setup+flotteur}(b)). It can thus enclose a tunable
volume of air, which enables to tune its apparent density, chosen for the
obstacle to float at the surface of the solution without applying an
undesirable vertical force on the top plate. In the presence of the foam, the
obstacle is in contact with the top plate through a capillary bridge, avoiding
solid friction. We check for each experiment that the obstacle is not stuck:
its position fluctuates under the slight flow heterogeneities, and results
presented below average the position of the obstacle over 50 successive images
with an interval of two seconds. Viscous friction cannot be eliminated, but it
only influences transients, which are not considered in this paper: each
measurement is performed in a stationary regime. Reversibility and
reproducibility tests give an upper bound for the force measurement errors: 0.2
mN, to be compared to the typical forces, of the order of 5 mN.

\subsection{Dissipation measurements}

The foam flowing in the channel experiences viscous friction, because of the
velocity gradients between the bubbles, the coverslip and the subphase, and it
exhibits energy dissipation through a pressure drop. If the channel remains
horizontal, the thickness of the foam thus decreases along the channel, because
its bottom is in contact with the subphase subject to hydrostatic pressure, and
the foam can even run over the tank at its open end. We overcome this
difficulty by tilting the whole setup thanks to a screw, so that the foam
recovers constant thickness along the channel. Furthermore, the level
difference $h$ one has to impose between the two ends of the channel provides a
simple measurement of the pressure drop $\Delta P$ of the foam through the
hydrostatic pressure in the subphase: $\Delta P = \rho gh$, where $\rho$ is the
volumetric mass of the solution and $g=9.8$ m$\cdot$s$^{-2}$ the gravity
acceleration. The thickness of the foam is measured at both ends of the channel
by eye, thanks to a graduated rule placed on the side of the channel. When
these measurements are done carefully, differences of thickness of $\Delta h =
0.1$ mm are detected. This yields a precision much better than the typical
pressure drop, which correspond to level differences of order 2 to 15 mm.

We have checked that when the thickness of the foam is equal at both ends, it
also remains constant along the channel, which means that the rate of
dissipation per unit length is also constant. Instead of the pressure drop
$\Delta P$, we will therefore deal with the pressure gradient $\nabla P =
\Delta P/L$, where $L=110$ cm is the length of the channel. We have also
checked that the pressure gradient does not depend on the presence of the
obstacle, measuring the same pressure drop with the three different obstacles
studied in this paper and without obstacle (data not shown). This enables to
consider the pressure gradient as the relevant parameter to quantify the
dissipation of the foam flowing in the channel.

\subsection{\label{ControlParameters}Control parameters}

A first control parameter is the nitrogen flow rate $Q$, which is adjusted
using an electronic controller (Brooks Instrument B.V.) driven by a home-made
software. The range of available flow rate runs on more that three decades,
from 1 to 2,000 mL/min, with a precision of 0.1 mL/min. Another control
parameter is the bubble volume. It is indirectly determined by measuring the
surface density of bubbles against the coverslip thanks to image analysis,
using NIH Image software. Since the foam thickness is kept equal to the initial
3.50 mm gap between the surface of the solution and the coverslip, there is a
univoque relation between the bubble volume and the mean surface density.
Instead of this surface density, we will refer throughout this paper to its
invert, that we call bubble area. This parameter differs slightly from the
bubble area one can measure directly on an image, because it includes the water
contained in the films and Plateau borders surrounding bubbles. For a given
injector, the bubble volume increases with the gas flow rate. To control these
two parameters separately, we blow the gas through one to six tubes (or
needles) of same diameter simultaneously, keeping constant the flow rate per
tube, hence the bubble volume. Furthermore, the diameter of these injectors can
be varied, which changes the flow rate per tube for the same bubble area;
hence, for a given bubble volume, typically ten different values of flow rate
are allowed. We always produce monodisperse foams: the bubble area disorder,
measured as the ratio of the standard deviation with the mean value of the
bubble area distribution, is less than 5\%. Six different bubble areas were
used: 12.1, 16.0, 20.0, 25.7, 31.7 and 39.3 mm$^2$, with a relative precision
of 3\%. The study of smaller bubbles would be problematic, since a transition
from bubble monolayer to multilayer occurs at low horizontal area/height ratio
\cite{Cox2002}. At the other extremity, it would be difficult to make a
monodisperse foam with larger bubbles.

Another tunable parameter is the viscosity of the solution. We control it by
adding glycerol to the initial soap solution. We have used five different
solutions, with 0, 20, 30, 40 and 50\% glycerol in mass. The respective
kinematic viscosities $\nu$, measured with a capillary viscometer
(Schott-Ger\"ate) at room temperature, are equal to 1.06, 1.6, 2.3, 3.8 and 9.3
mm$^2$/s. The variation of viscosity due to the variation of room temperature
is lower than 4\%.

\begin{figure}
\begin{center}
\includegraphics[width=6cm]{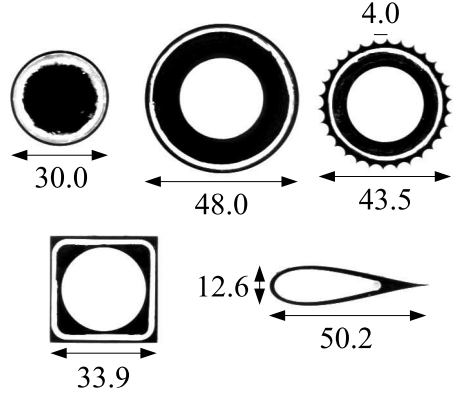}
\caption{\label{Obstacles} Top views of the three
obstacles.}\end{center}
\end{figure}

Different obstacles have been used. The basic obstacle, whose density is
tunable, is a cylinder of diameter 30 mm on which additional profiles can be
fixed. We have studied three different obstacles: two cylinders, of diameter 30
and 48 mm, and a cogwheel of diameter 43.5 mm, with circular cogs of diameter 4
mm (Fig. \ref{Obstacles}). For each obstacle, the apparent density is adjusted
as described above to avoid solid friction. As said previously, the presence of
the obstacle influences the measurements of drag, but not those of dissipation.

As a final remark about control parameters, for given solution viscosity, area
and obstacle, various flow rates are available (from 5 to 13 in the following
data), with greatest flow rate at least 20 times greater than the lowest one.

\section{Drag measurements}

\subsection{\label{SectionViscosity}Influence of solution viscosity}

We study the variation of the drag \emph{versus} the flow rate and the solution
viscosity, for the five different viscosities indicated in subsection
\ref{ControlParameters}. All these measurements are performed at a fixed bubble
area of 20 mm$^2$, and we use a circular obstacle of diameter 30 mm.

\begin{figure}
\begin{center}
\includegraphics[width=8cm]{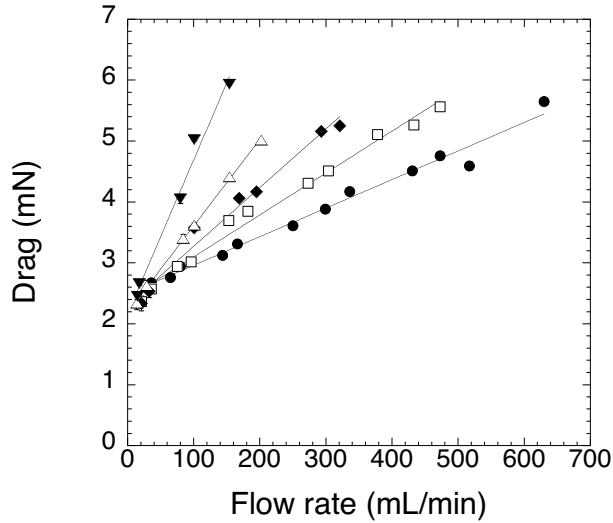}
\caption{\label{Drag (viscosity)} Drag \emph{versus} flow rate, for solution
viscosity equal to 1.06 ($\bullet$), 1.6 ($\square$), 2.3 ($\blacklozenge$),
3.8 ($\vartriangle$) and 9.3 mm$^2$/s ($\blacktriangledown$). The straight
lines are linear fits of the data. The bubble area is 20 mm$^2$ and the
obstacle is a circle of diameter 30 mm.}\end{center}
\end{figure}

\begin{figure}
\begin{center}
\includegraphics[width=7cm]{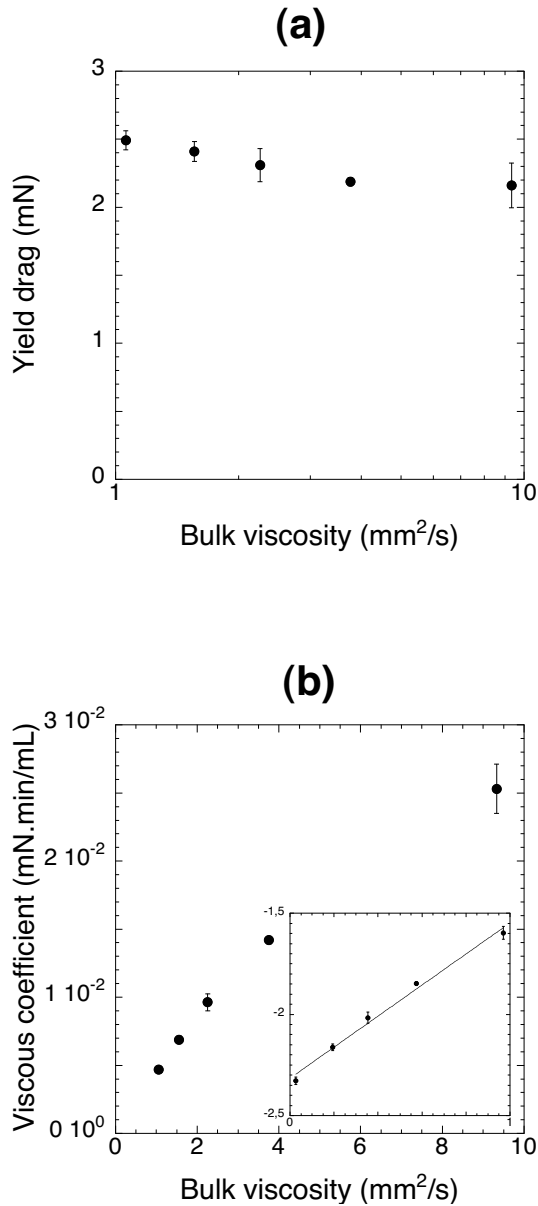}
\caption{\label{Viscosity} Results from fits to Fig. \ref{Drag (viscosity)}.
(a) Yield drag \emph{versus} the solution viscosity (semi-logarithmic scale),
and (b) viscous coefficient \emph{versus} the solution viscosity (linear
scale). Insert: log-log plot. The straight line is the linear fit: its slope is
$0.77 \pm 0.05$.}\end{center}
\end{figure}

We observe two general features (Fig. \ref{Drag (viscosity)}), independent of
the value of the solution viscosity: the drag does not tend to zero at low flow
rate, and it increases with flow rate. The first observation is a signature of
the solid-like properties of the foam. The second feature is related to the
fluid-like properties of the foam. The data are well fitted by a linear law
(Fig. \ref{Viscosity}):
\begin{equation}\label{Linear fit}
  F=F_0+mQ.
\end{equation}
We call $F_0$ the yield drag, as a reference to the yield properties of the
foam, and the slope $m$ the viscous coefficient, since we can dimensionally
deduce from $m$ an effective viscosity $\mu$ for the foam: $\mu\approx mS/R$,
where $S$ is the cross-section of the foam, and $R$ is the typical size of the
obstacle. Yield drag \emph{versus} solution viscosity is plot on Fig.
\ref{Viscosity}(a), and viscous coefficient \emph{versus} solution viscosity on
Fig. \ref{Viscosity}(b).

Fig. \ref{Viscosity}(a) shows that the yield drag is essentially independant of
the solution viscosity. This was expected, because yield drag is only related
to the yield properties of the foam, which depend on surface tension and bubble
size \cite{Princen1983}. The slight decrease with the solution viscosity is
likely due to a slight decrease of surface tension with the glycerol rate in
the solution, as quantified for aqueous mixtures of glycerol without
surfactant, whose surface tension decreases by 7\% from pure water to a mixture
with half glycerol in mass \cite{Handbook2003}.

Fig. \ref{Viscosity}(b) shows that the viscous coefficient increases with the
solution viscosity. The data can be fitted by a power law (insert of Fig.
\ref{Viscosity}(b)), that yields the following dependency of viscous
coefficient on solution viscosity: $m \propto \nu^{0.77\pm 0.05}$.

\subsection{\label{SectionArea}Influence of bubble area}

We now turn to the study of drag versus flow rate and bubble area. All the
measurements are done without adding glycerol in the solution, at a constant
viscosity of 1.06 mm$^2$/s. The obstacle is a cylinder of radius 30 mm. We
study the six bubble areas indicated in subsection \ref{ControlParameters},
from 12.1 mm$^2$ to 39.3 mm$^2$.

\begin{figure}
\begin{center}
\includegraphics[width=8cm]{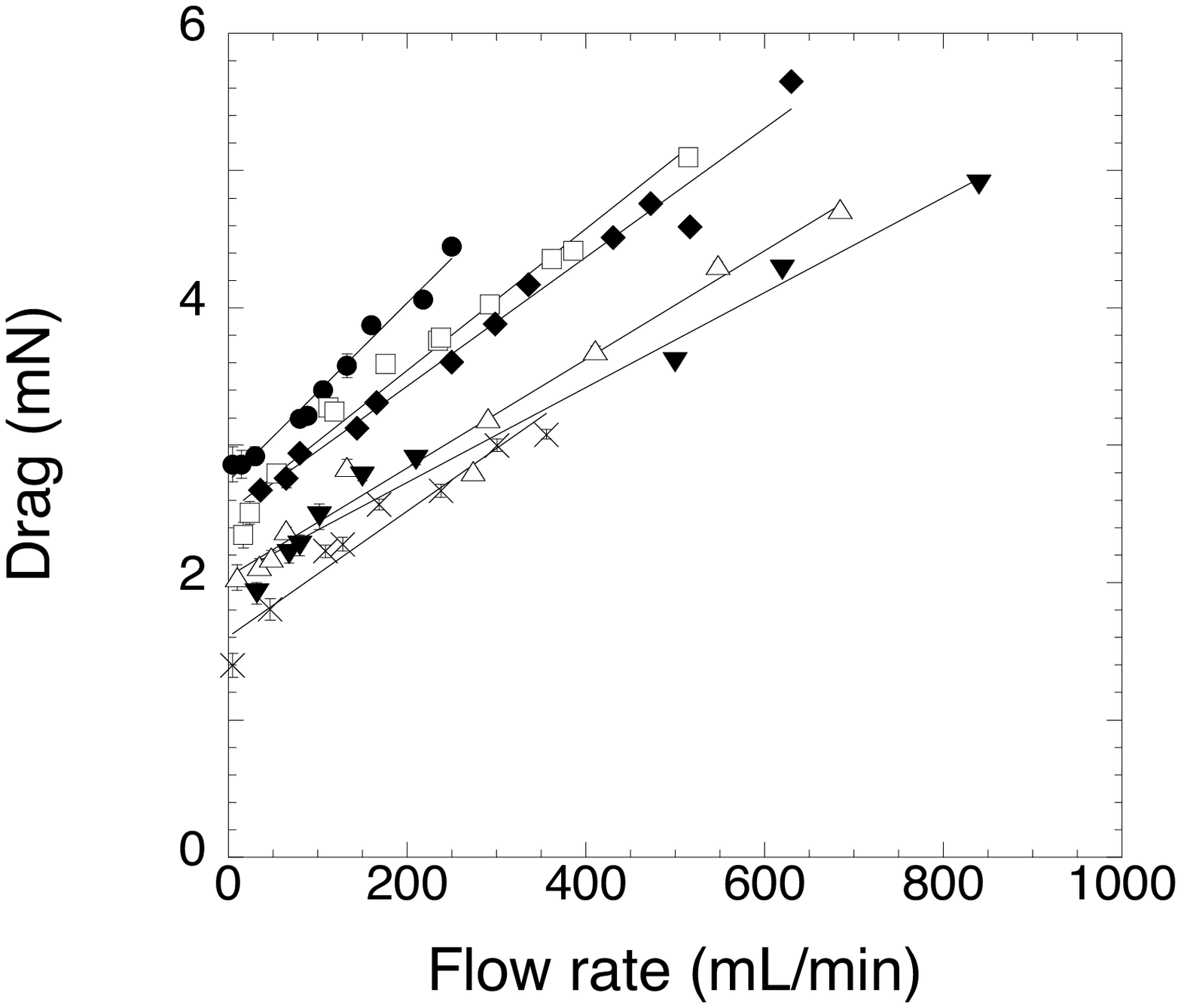}
\caption{\label{Drag (area)} Drag \emph{versus} flow rate, for bubble area
equal to 12.1 ($\bullet$), 16.0 ($\square$), 20.0 ($\blacklozenge$), 25.7
($\vartriangle$), 31.7 ($\blacktriangledown$) and 39.3 mm$^2$ ($\times$). The
straight lines are linear fits of the data. The solution viscosity is 1.06
mm$^2$/s and the obstacle is a circle of diameter 30 mm.}\end{center}
\end{figure}

\begin{figure}
\begin{center}
\includegraphics[width=7cm]{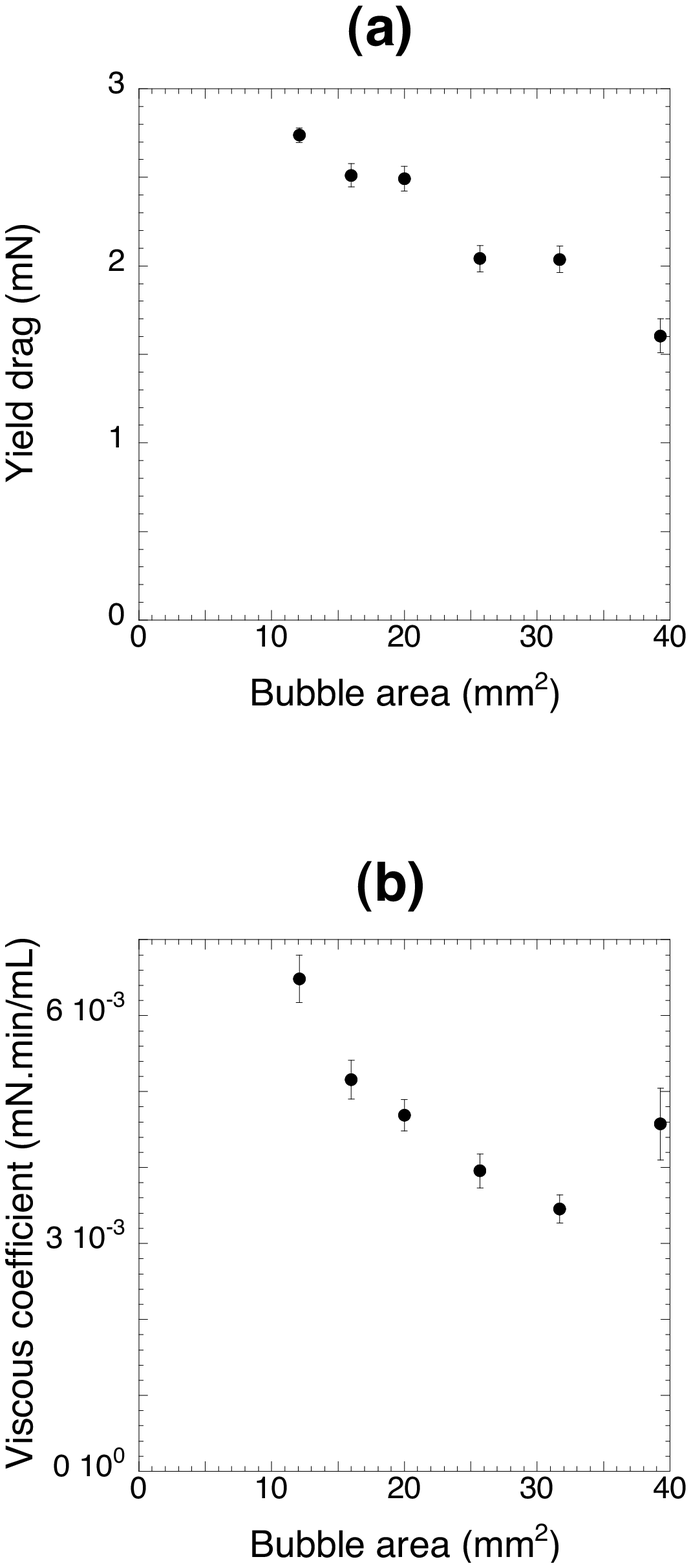}
\caption{\label{Area} Results from fits to Fig. \ref{Drag (area)}. (a) Yield
drag, (b) viscous coefficient \emph{versus} bubble area.}\end{center}
\end{figure}

We find again the signature of the viscoplastic properties of the foam (Fig.
\ref{Drag (area)}), with a non-zero yield drag and an increase of drag
\emph{versus} flow rate. Performing the linear fit (\ref{Linear fit}), we get
the yield drag and the viscous coefficient, plotted \emph{versus} bubble area
in Fig. \ref{Area}.

Fig. \ref{Area}(a) evidences that the yield drag is a decreasing function of
the bubble area. This is coherent with the fact that both quantities used to
describe the solid properties of the foam, its shear modulus and yield stress,
are also decreasing functions of the bubble size
\cite{Princen1985,Mason1995,Mason1996}. Fig. \ref{Area}(b) shows that the
viscous coefficient is also a decreasing function of bubble area, except for
the last point. Further analysis of these data is somewhat complex, and will be
discussed in more detail in subsection \ref{DiscussionArea}.

\subsection{\label{SectionObstacle}Influence of obstacle size and boundary conditions}

We now study a third control parameter: the obstacle geometry. As indicated in
subsection \ref{ControlParameters}, we use two cylinders of different radius
and a cogwheel. This enables to study the influence of the size of the obstacle
and of the boundary conditions: the foam slips along the smooth cylinders
whereas the first layer of bubbles around the cogwheel is anchored in the cogs.
As in the previous subsection, the solution of viscosity of 1.06 mm$^2$/s is
used. A bubble area of 16.0 mm$^2$ was chosen in order to adapt the bubble size
to the cog diameter, for the bubbles to be correctly trapped in the cogs.

\begin{figure}
\begin{center}
\includegraphics[width=8cm]{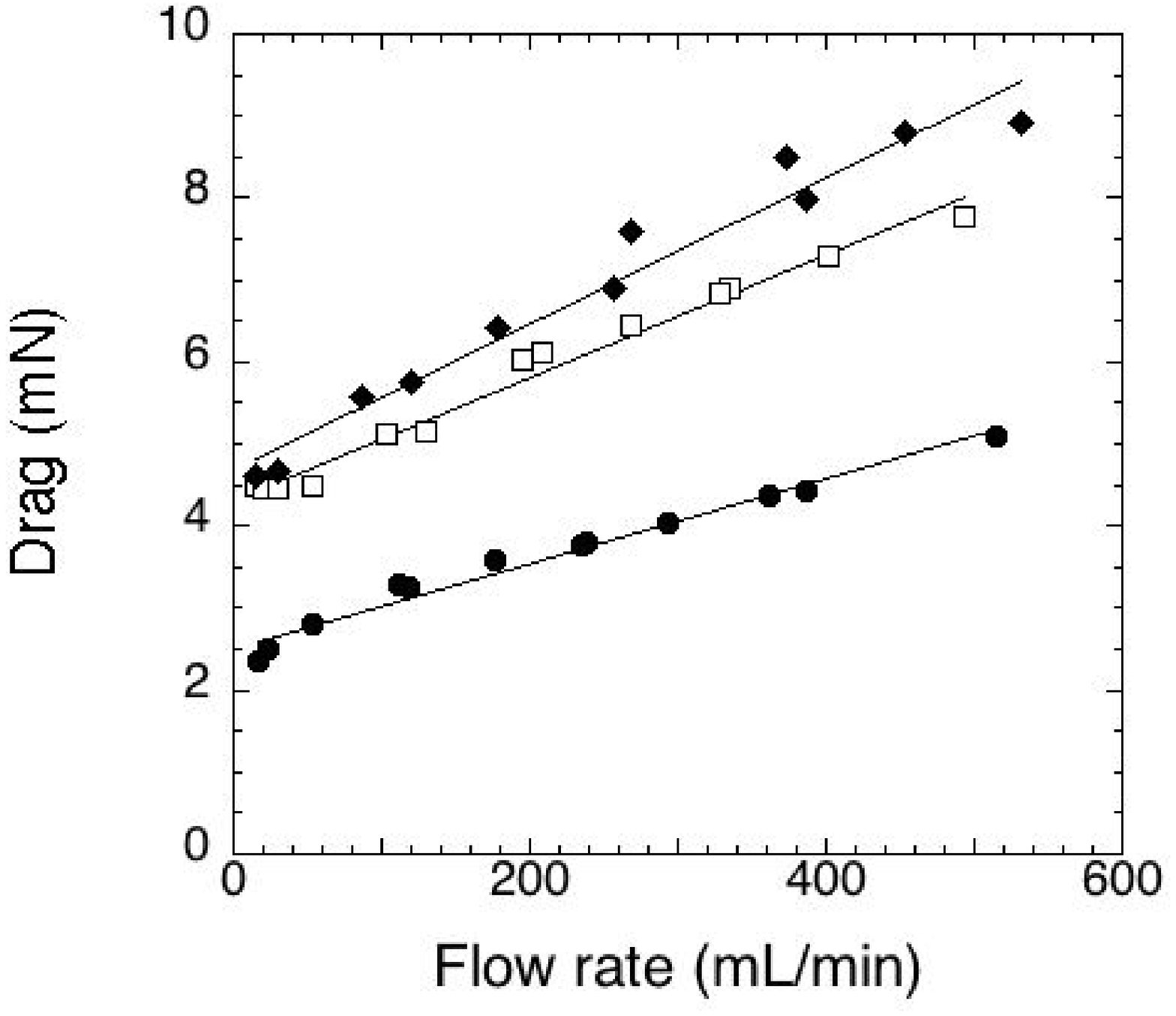}
\caption{\label{Drag (obstacle)} Drag \emph{versus} flow rate, for
the following obstacles: cylinder of diameter 30 mm ($\bullet$),
cogwheel of diameter 43.5 mm with cylindrical teeth of diameter 4
mm ($\square$), and cylinder of diameter 48 mm
($\blacklozenge$).}\end{center}
\end{figure}

\begin{figure}
\begin{center}
\includegraphics[width=10cm]{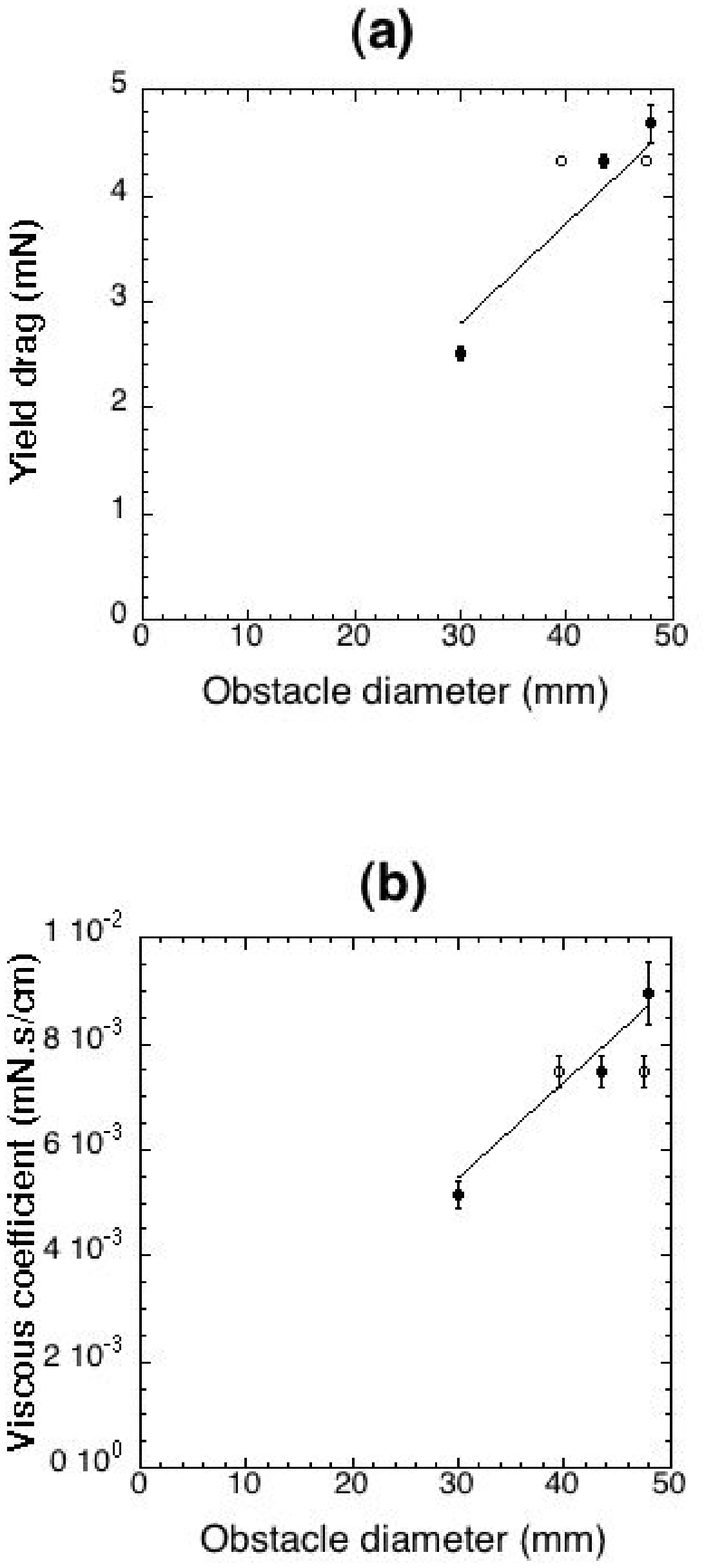}
\caption{\label{Obstacle} (a) Yield drag, and (b) viscous coefficient
\emph{versus} the obstacle diameter. The cogwheel is represented by three
points, which correspond to the three possible choices for its diameter: inner
diameter of 39.5 mm (left $\circ$), mean diameter of 43.5 mm($\bullet$), and
outer diameter of 47.5 mm (right $\circ$). The straight line is a linear fit to
the two cylinders data passing through zero.}\end{center}
\end{figure}

As expected, the drag increases with the size of the obstacle (Fig. \ref{Drag
(obstacle)}). The results of the linear fit (\ref{Linear fit}) are displayed in
Fig. \ref{Obstacle}(a) and \ref{Obstacle}(b) \emph{versus} obstacle diameter.
There are three possible choices for the cogwheel diameter: a mean diameter of
43.5 mm, which would be the diameter of the obstacle without cogs, an inner
diameter of 39.5 mm if the cogs are excluded, and an outer diameter of 47.5 mm
if the bubbles trapped in cogs are included. Hence, three points are
represented for inner, mean and outer diameter of the cogwheel in Fig.
\ref{Obstacle}.

Fig. \ref{Obstacles} shows that the increase of drag with obstacle diameter is
close to linear, and we fit the data by a linear law passing through zero. This
is a way to study the influence of boundary conditions, since it enables to
consider the cogwheel as an effective obstacle, whose effective diameter is
given by the fitting line for the values of yield drag and viscous coefficient
of the cogwheel. This effective diameter is to compare to the three possible
choices described above for the diameter of the cogwheel. Concerning the yield
drag, Fig. \ref{Obstacle}(a) shows that the effective diameter is close to the
outer one, whereas for the viscous coefficient, Fig. \ref{Obstacle}(b) shows
that it is close to the inner one. This difference between the behaviour of the
cogwheel for yield drag and viscous coefficient is discussed in the next
section.

\section{\label{Dissipation}Dissipation measurements}

For each experiment, the drag and the pressure gradient are simultaneously and
independently measured. We thus present systematic measurements for the
pressure gradient like for the drag, studying the same control parameters
except the obstacle, whose presence do not change the results, as mentioned
above.

\subsection{Influence of solution viscosity}

We study the variation of the pressure gradient \emph{versus} the flow rate and
the solution viscosity, for the five different viscosities indicated in
subsection \ref{ControlParameters}. All these measurements are performed at a
fixed bubble area of 20 mm$^2$.

\begin{figure}
\begin{center}
\includegraphics[width=10cm]{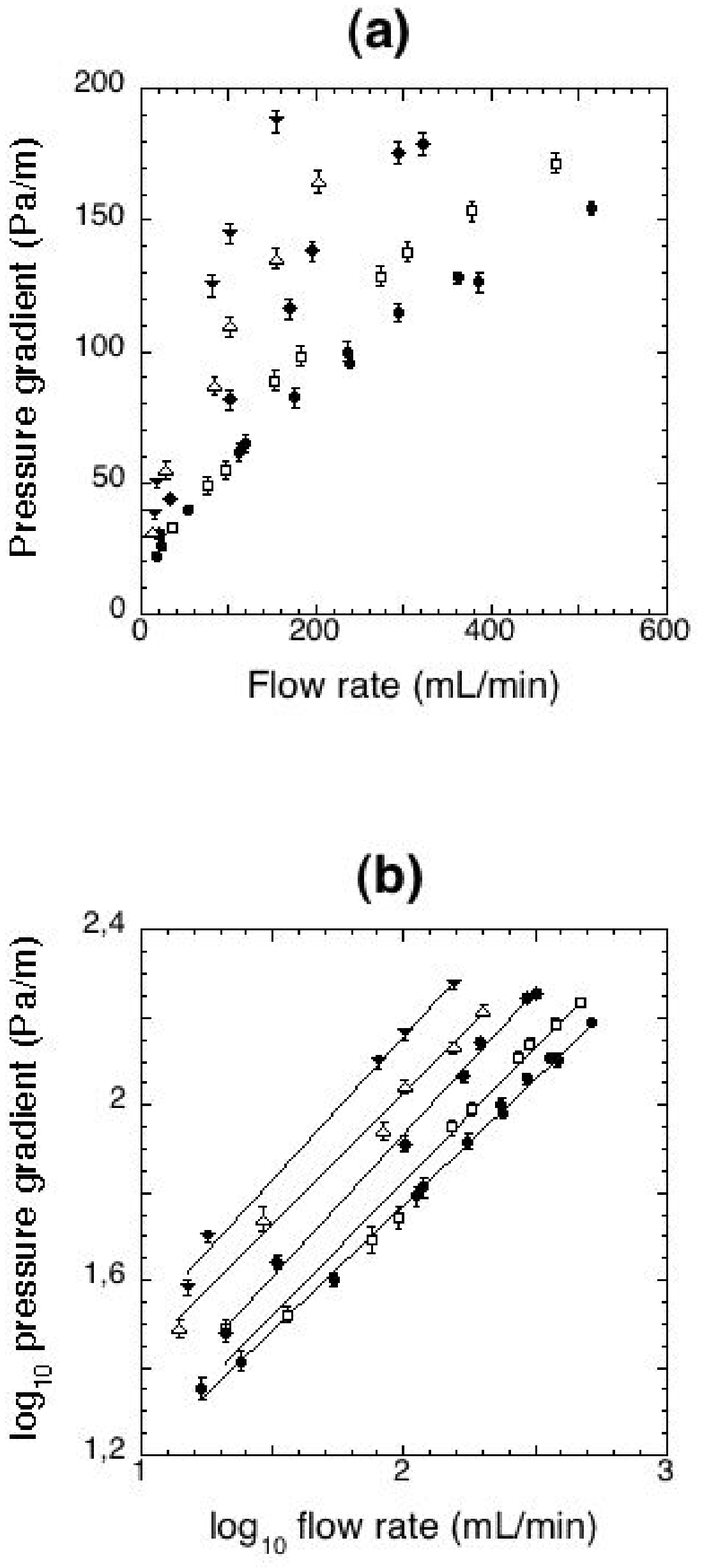}
\caption{\label{PrGradViscosity} (a) Pressure gradient \emph{versus} flow rate,
for solution viscosity equal to 1.06 ($\bullet$), 1.6 ($\square$), 2.3
($\blacklozenge$), 3.8 ($\vartriangle$) and 9.3 mm$^2\cdot$s$^{-1}$
($\blacktriangledown$). (b) Log-log plot of the same data.}\end{center}
\end{figure}

\begin{figure}
\begin{center}
\includegraphics[width=7cm]{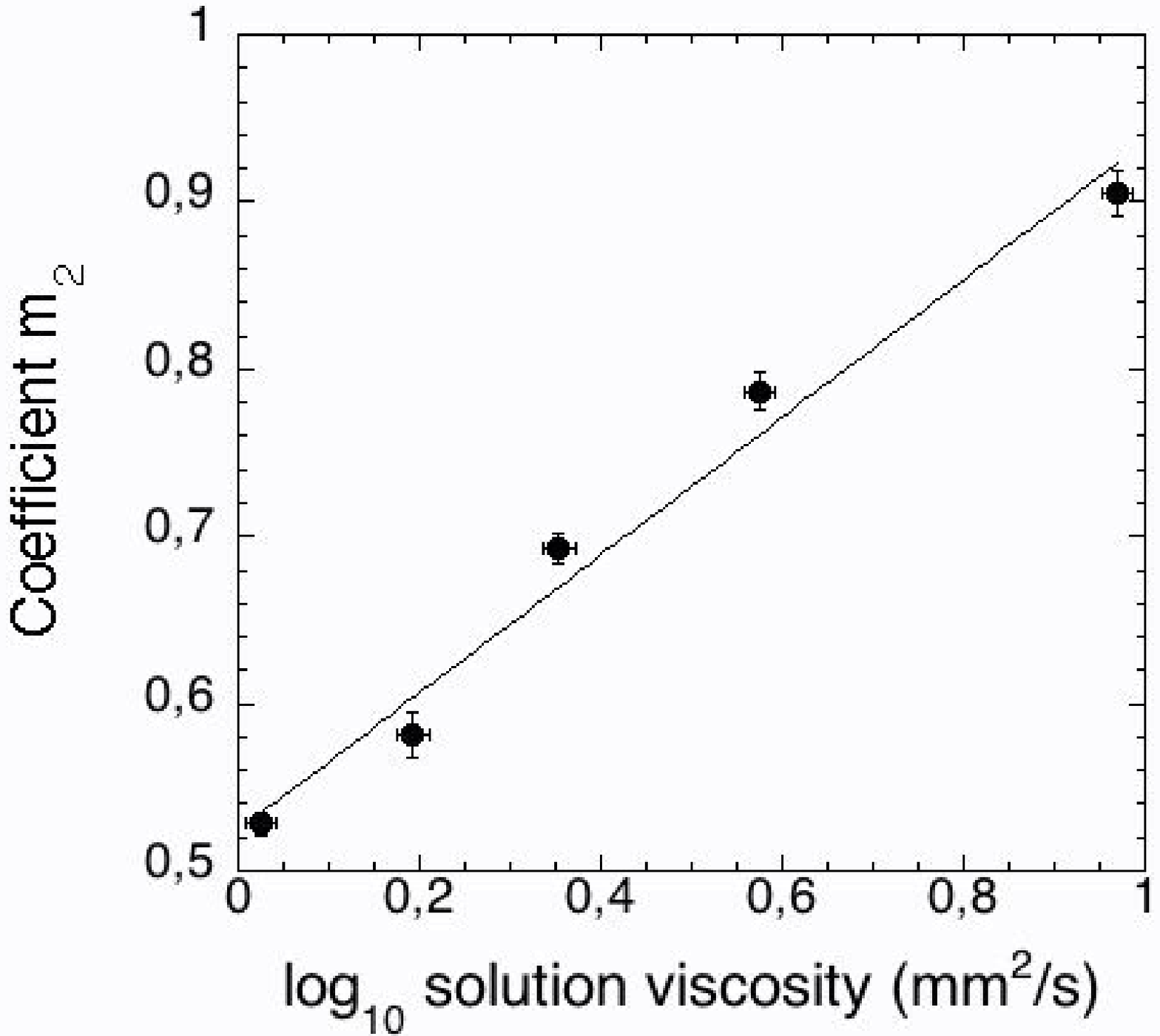}
\caption{\label{Coeff_m2_pr_grad(visc)} Coefficient $m_2$ of the
fit $\log \nabla P = 0.62\log Q + m_2$ \emph{versus} the logarithm
of the solution viscosity. The straight line is a linear fit of
the data.}\end{center}
\end{figure}

We observe that the pressure gradient increases with both the flow rate and the
solution viscosity (Fig. \ref{PrGradViscosity}(a)). These tendencies are
quantified by the log-log plot (Fig. \ref{PrGradViscosity}(b)). For each
solution viscosity, the data are well linearly fitted, indicating that the
pressure gradient depends on the flow rate with a power-law dependence.
Furthermore, all fitting lines are nearly parallel; this exponent is thus
independent of the solution viscosity, and its value obtained by averaging over
the five solution viscosities equals $0.62\pm 0.03$. To quantify the dependency
of the pressure gradient on the solution viscosity, we thus use the following
fit: $\log \nabla P = 0.62\log Q + m_2$, and plot the coefficient $m_2$, as a
fonction of the logarithm of the solution viscosity in Fig.
\ref{Coeff_m2_pr_grad(visc)}. The data are again well fitted by a linear law,
indicating another power-law dependence of the pressure gradient, on the
solution viscosity, with an exponent equal to $0.41\pm 0.04$.

\subsection{Influence of bubble area}

We now present the study of pressure gradient versus flow rate and bubble area.
All the measurements are done without adding glycerol in the solution, at a
constant viscosity of 1.06 mm$^2$/s. We study the six bubble areas indicated in
subsection \ref{ControlParameters}.

\begin{figure}
\begin{center}
\includegraphics[width=10cm]{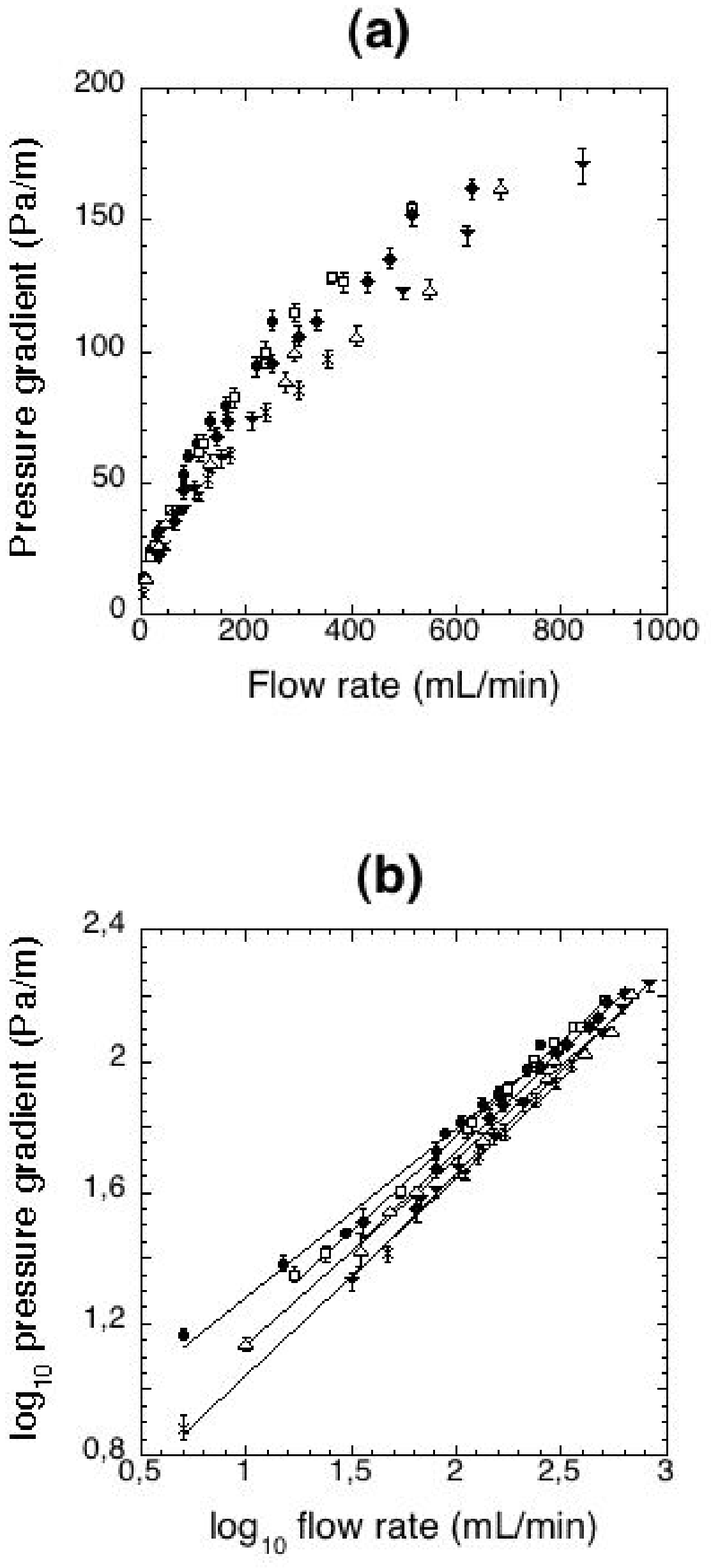}
\caption{\label{PrGradArea} (a) Pressure gradient \emph{versus} flow rate, for
bubble area equal to 12.1 ($\bullet$), 16.0 ($\square$), 20.0
($\blacklozenge$), 25.7 ($\vartriangle$), 31.7 ($\blacktriangledown$) and 39.3
mm$^2$ ($\times$). (b) Log-log plot of the same data.}\end{center}
\end{figure}

We observe that the pressure gradient increases again with the flow rate, and
decreases with the bubble area (Fig. \ref{PrGradArea}(a)). The log-log plot is
displayed in Fig. \ref{PrGradArea}(b). The relative variation of bubble area is
smaller than the one of solution viscosity, but for each bubble area, the data
are again well linearly fitted with fitting lines nearly parallel, yielding an
exponent of $0.58\pm 0.04$ for the power-law dependance of the pressure
gradient on the flow rate. This exponent is compatible with the one obtained in
the previous subsection, to within the experimental errors. To quantify the
dependency of the pressure gradient on the bubble area, we use the fit: $\log
\nabla P = 0.58\log Q + m_2$, and plot the coefficient $m_2$, as a fonction of
the logarithm of the bubble area in Fig. \ref{Coeff_m2_pr_grad(visc)}. The data
are remarkably linearly fitted, indicating a third power-law dependence of the
pressure gradient, on the bubble area, with an exponent equal to $-0.33\pm
0.01$.

\begin{figure}
\begin{center}
\includegraphics[width=7cm]{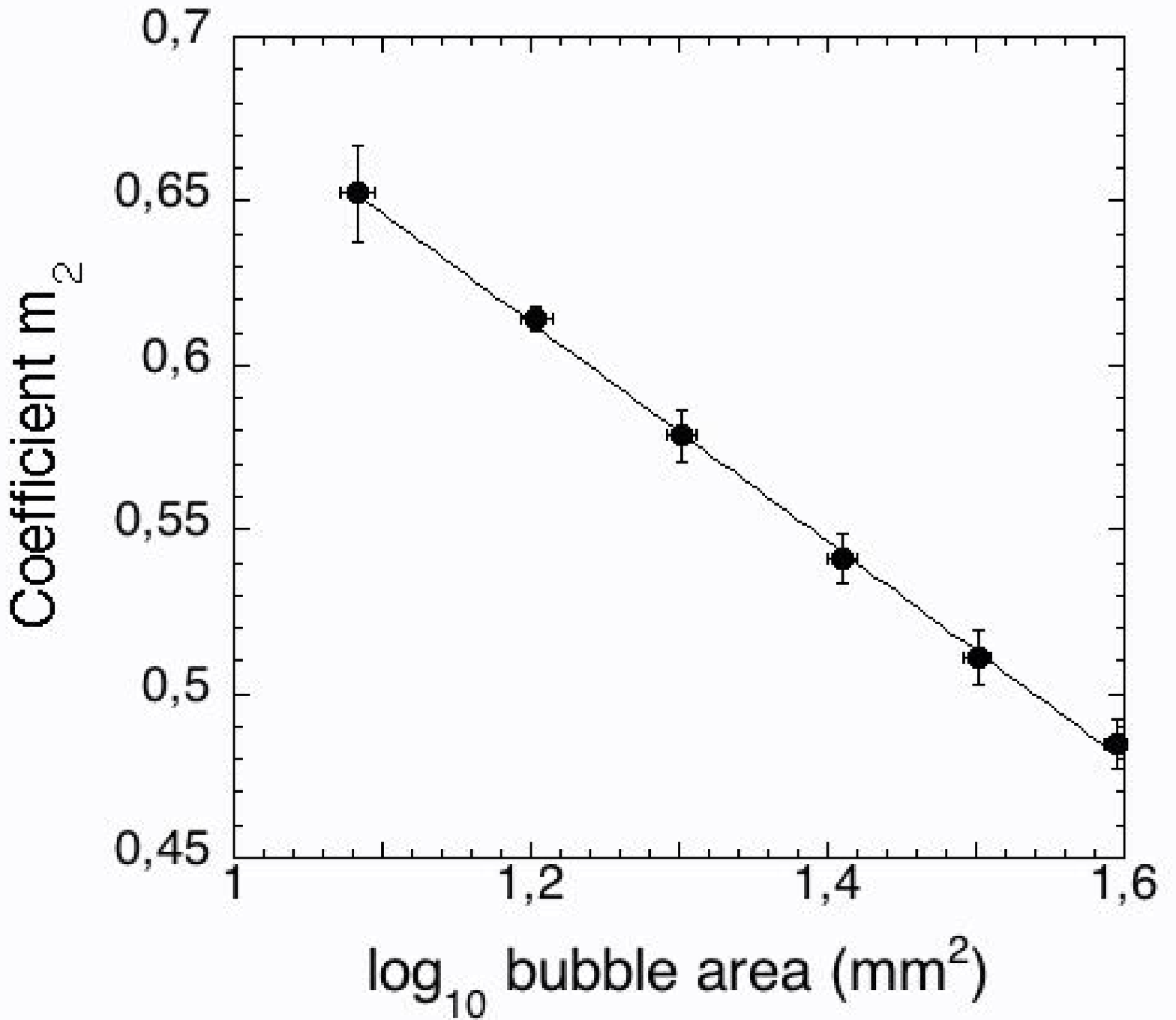}
\caption{\label{Coeff_m2_pr_grad(area)} Coefficient $m_2$ of the
fit $\log \nabla P = 0.58\log Q + m_2$ \emph{versus} the logarithm
of the bubble area. The straight line is a linear fit of the
data.}\end{center}
\end{figure}

To conclude this section about the dissipation measurements, we have shown that
the pressure gradient has a power-law dependance on the three studied control
parameters: flow rate, solution viscosity and bubble area, with three
independent exponents: $\nabla P \propto Q^\alpha \nu^\beta A^\delta$. However,
the values we got for $\beta$ and $\delta$ depended on two different chosen
values for $\alpha$, respectively 0.62 and 0.58, and they are therefore biased.
Unbiased values are obtained by direct fit of the whole data (89 independent
experiments) by the triple power-law above, and we finally get the following
scaling for the pressure gradient:
$$ \nabla P (\mathrm{Pa/m}) = (10.0\pm 0.4) \left( \frac{Q}{Q_0} \right)^{0.59\pm 0.01}
\left( \frac{\nu}{\nu_0} \right)^{0.46\pm 0.02} \left( \frac{A}{A_0}
\right)^{-0.34\pm 0.03} , $$ where $Q_0 = 1$ mL/min, $\nu_0 = 1$ mm$^2$/s and
$A_0 = 1$ mm$^2$. Note that as expected, the precision of the exponent is
better for parameters running on a wider range.

\section{Discussion}

\subsection{Influence of solution viscosity}

Our measurements of drag and pressure gradient \emph{versus}
viscosity $\nu$ and flow rate $Q$ yield the following scalings:
\begin{equation}\label{F(Q,nu)}
  F(Q,\nu) = F_0 + \mathrm{const}\times \nu^{0.77\pm 0.05}Q  ,
\end{equation}
and:
\begin{equation}\label{nablaP(Q,nu)}
  \nabla P(Q,\nu) = \mathrm{const}\times \nu^{0.46\pm 0.02}Q^{0.59\pm 0.01}
  .
\end{equation}
 (see subsection \ref{SectionViscosity} and section \ref{Dissipation}). To our knowledge,
this is the first time than such scalings are proposed to quantify the
dynamical regime of flowing foams in channels. Up to now, the dynamic regime of
flowing foam has been mainly investigated through the study of pressure drop of
foam confined in capillaries (see Cantat et al. \cite{Cantat2004} and
references therein). Since the early work of Bretherton \cite{Bretherton1961},
who studied the friction between an infinitely long bubble and a solid wall,
all these studies emphasize the role of the capillary number $\mathrm{Ca}=\eta
V/\gamma$, where $\eta$ is the dynamic viscosity of the solution, $\gamma$ its
surface tension and $V$ the velocity of the flowing foam. In the frame of our
study, the capillary number is proportional to the product $\nu Q$. It appears
from our scalings (\ref{F(Q,nu)}) and (\ref{nablaP(Q,nu)}) that such a number
is not sufficient to describe the dynamic regime of a flowing foam, because the
exponents for viscosity and flow rate differ significantly. Bretherton's theory
is therefore not sufficient to explain our measurements: additional physical
ingredients are involved, like detailed bubble shape and interfacial rheology
(surface elasticity and viscosity). This has not been investigated yet.
However, it is worth noting that from the scalings (\ref{F(Q,nu)}) and
(\ref{nablaP(Q,nu)}), the quantity $[F(Q,\nu)-F_0]^{0.59}/\nabla P(Q,\nu)$ does
not significantly depend neither on flow rate, nor on solution viscosity. This
confirms that both the dissipation and the velocity-dependent part of the drag
are generated by a common mechanism, the viscous friction between bubbles and
solid boundaries.

Furthermore, the scaling (\ref{nablaP(Q,nu)}) shows that the exponent of the
pressure gradient significantly departs from $2/3$, which is the exponent
expected for tangentially perfectly mobile interfaces (implicit assumption in
Bretherton's theory \cite{Bretherton1961}), as well as from $1/2$, predicted
value for rigid interfaces \cite{Denkov2004}. This probably means that with the
used surfactants, the behaviour of the interfaces lies between these two
extreme cases.

\subsection{\label{DiscussionArea}Influence of bubble area}

The measurements of drag \emph{versus} bubble area and flow rate, in subsection
\ref{SectionArea}, show that both yield drag and viscous coefficient are
decreasing functions of the bubble area. This is expected for the yield drag,
since the foam shear modulus and yield stress are decreasing functions of the
bubble size \cite{Princen1985,Mason1995,Mason1996}. However, the yield stress
of the foam is not the only contribution to the yield drag, and preliminary
simulations of our experiments \cite{Cox} show that the resultant of the
pressure of bubbles in contact with the obstacle is a second non-negligible
contribution to the yield drag, and that these two contributions act in the
same sense. Since we are not able to measure the pressure in the bubbles, we
cannot quantify the pressure contribution in our experiments, so the
interpretation of the evolution of yield drag \emph{versus} bubble area (Fig.
\ref{Area}(a)) is difficult and requires further numerical simulations.

Another major difficulty for quantitative interpretation arises from the
variation of fluid fraction with bubble area. In our setup, the monolayer of
bubbles is in contact with a reservoir of water, and the amount of water in the
Plateau borders and films between bubbles is freely chosen by the system. For
instance, we have observed that little bubbles are more closely packed than big
ones. Therefore, the mean fluid fraction should vary with bubble area.
Furthermore, local effects such as dilatancy \cite{Weaire2003} could increase
the fluid fraction near the obstacle, because of the shear experienced by the
foam in this zone. This complicates the interpretation of the evolution of
yield drag with bubble area, because many studies have shown that rheological
properties of foams and emulsions strongly depend on fluid fraction
\cite{Mason1995,Mason1996,Saint-Jalmes1999}.

Despite the complications due to the effect of fluid fraction, we can propose a
qualitative argument to explain why the viscous coefficient decreases with the
bubble area, based on the dissipation model of Cantat and coworkers
\cite{Cantat2004}. These authors state that dissipation in flowing foam is
localised in the Plateau borders between bubbles and walls. Hence, dissipation
should increase with the number of bubbles surrounding the obstacle, and
therefore the viscous coefficient should decrease with the bubble area, which
is actually seen in Fig. \ref{Area}(b). The scaling of the pressure gradient
with the bubble area, $\nabla P \propto A^{-0.34\pm 0.03}$ (see section
\ref{Dissipation}), is also compatible with this model. However, we should
expect the pressure gradient to be proportional to the total length of the
Plateau borders per unit area of foam. Since the foams are monodisperse in our
experiments, we should thus expect a scaling $\nabla P \propto 1/\sqrt{A}$,
which differs from ours. Actually, this scaling applies for sharp Plateau
borders, matching a flat thin film between a bubble and a wall and another one
between two bubbles. Such an idealisation does not apply to our experiments,
since the bubble shape is essentially curved, hence Plateau borders are much
smoother. Furthermore, the shape of the bubbles strongly depends on their
volume, because of the influence of buoyancy. Note also that this model does
not capture the increase of viscous coefficient observed for the bubble area of
39.3 mm$^2$.

As an additional remark, friction in the foam should strongly depend on the
boundary conditions at the interfaces between films and bubbles, hence the
viscous coefficient probably changes with the surface rheology. It would thus
be interesting to investigate the influence of the surfactant used on the drag
measurements.

\subsection{Influence of the obstacle geometry}

The measurements of drag for different obstacles, presented in subsection
\ref{SectionObstacle}, show that both the yield drag and the viscous
coefficient increase linearly with the diameter of the obstacle, and that the
effect of the boundary conditions on the obstacle is not marked. We now discuss
these two observations.

Our measurements of drag around circular obstacles are to compare to the
theoretical value of the drag exerted by a Newtonian fluid of dynamic viscosity
$\eta$, flowing at velocity $V$, on a cylindrical obstacle of radius $R$ in a
channel of width $2H$ \cite{Faxen1946}:
\begin{equation} \label{Faxen}
F \simeq \frac{4\pi\eta RV}{\ln H/R - 0.91}.
\end{equation}
It is worth noting that this law does not predict proportionality between the
drag and the obstacle diameter when $R/H$ ranges from $0.3$ to $0.5$, contrary
to our observations. This indicates again that the elastic properties of the
foam are significant, even when we consider only the evolution of viscous
coefficient \emph{versus} obstacle diameter (Fig. \ref{Obstacle}(b)). This is
to compare to the simulations of Mitsoulis and coworkers
\cite{Zisis2002,Mitsoulis2004}, who computed the drag exerted by a flowing
Bingham plastic on a cylinder in the same geometry than ours, for different
values of obstacle diameters. A Bingham plastic is characterized by its yield
stress $\tau_y$ and its plastic viscosity $\mu$, and it follows the
constitutive equation: $\tau = \tau_y + \mu\dot{\gamma}$ for $|\tau| > \tau_y$,
and $\dot{\gamma} = 0$ for $|\tau| < \tau_y$, where $\tau$ is the shear stress
and $\dot{\gamma}$ the applied strain. To summarize, Mitsoulis and coworkers
show that the drag exerted by a flowing Bingham plastic around a cylinder
strongly depends on the Bingham number $\mathrm{Bn}=2R\tau_y/\eta V$ comparing
elastic and viscous contribution: at a given Bingham number of order unity,
there is a crossover between a Newtonian-like behaviour of the drag (for
$\mathrm{Bn} \ll 1$) given by formula (\ref{Faxen}), and an elastic-like (for
$\mathrm{Bn} \gg 1$) where drag does not significantly depend on the velocity
and is roughly proportional to the cylinder diameter. Though modeling foam as a
Bingham plastic is an open debate, this work provides an interesting comparison
to our experimental measurements, for which we now evaluate the order of
magnitude of the Bingham number in our experiments. The yield stress for a foam
is of order \cite{Princen1983} $0.5\gamma/a$, with $\gamma=31$ mN/m the surface
tension and $a\approx \sqrt{16/(3^{3/2}/2)}\approx 2.5$ mm the typical length
of a bubble edge (we recall that the bubble area is 16.0 mm$^2$ in the
considered experiments, and compute $a$ for an hexagonal bubble), so $\tau_y
\approx 6$ Pa (to be rigorous, this overestimates the yield stress for a wet
foam). Furthermore, we can deduce from the value of the viscous coefficient ($m
= 5\times 10^{-6}$ N$\cdot$min/mL after Fig. \ref{Area}(b)) a rough value of
the plastic viscosity of the foam: dimensional analysis yields $\mu \approx
mS/R$ where $S$ is the cross-section of the foam, so Bingham number writes
$\mathrm{Bn} \approx 2R^2 \tau_y/mQ$. The typical value of flow rate in our
experiments is $10^2$ mL/min, hence the typical Bingham number equals
$\mathrm{Bn} \approx (2\times 0.015^2\times 6)/(5\times 10^{-6}\times 10^2)
\approx 5$. Though this is a very rough evaluation of the Bingham number, this
tends to confirm that in our range of flow rates, this parameter remains of
order unity, hence both elastic and fluid properties of the foam are involved
in the interaction with the obstacle to create the drag.

We have observed in subsection \ref{SectionObstacle} that the effective
diameter of the cogwheel is different if we consider the yield drag or the
viscous coefficient. For the yield drag, the cogwheel behaves like a large
cylinder, thus including the trapped bubbles. On the other hand, for the
viscous coefficient, the cogwheel behaves like a smaller cylinder. Actually,
the cogwheel and the trapped bubbles form a closed system during the
experiment: no rearrangement of the trapped bubbles occurs after all the teeth
have been filled with bubbles. So this system behaves as an effective obstacle,
but with an external boundary constituted of bubble edges, instead of a solid
boundary. This explains the difference observed between the yield drag and the
viscous coefficient: at low velocity, the foam feels the presence of the
effective obstacle, but at high velocity, the friction between the effective
obstacle and the surrounding flowing bubbles is lower than the friction between
a solid obstacle and its neighbouring flowing bubbles. To be more quantitative,
it would be interesting to study the influence of interfacial rheology on this
friction. Anyway, the measurements show that the influence of boundary
conditions is not dramatic, probably because it does not change much the
features of the flow beyond the first layer of bubbles.

\section{Conclusions}

This work provides the first detailed and systematic measurements
of the force exerted by a 2D flowing foam on an obstacle as a
function of various control parameters: flow rate, solution
viscosity, bubble volume and obstacle shape and size. All the data
show two contributions to the drag: a yield drag for flow rate
tending to zero, and a flow rate-dependant contribution. We have
shown that the yield drag is independent of the solution
viscosity, decreases with bubble volume and linearly increases
with the obstacle diameter. Fitting the flow rate-dependant
contribution by a linear law, we have shown that the slope (or
viscous coefficient) increases with the solution viscosity as a
power law with an exponent of $0.77\pm 0.05$; moreover, the
viscous coefficient globally decreases with the bubble volume and
linearly increases with the obstacle diameter. Furthermore, we
have pointed out that the effect of boundary conditions on the
obstacle is not striking.

This paper also presents a systematic study of the dissipation of
a 2D flowing foam in a channel as a function of the same control
parameters that for the drag. Dissipation is quantified by a
pressure gradient, which is independent of the presence of an
obstacle, and scales with the three parameters with a power-law
dependence: the exponents equal $0.59\pm 0.01$ for the flow rate,
$0.46\pm 0.02$ for the solution viscosity, and $-0.34\pm 0.03$ for
the bubble area. Though this scaling is quite accurate, it remains
difficult to interpret, both because of the complex bubble
geometry and because the interfacial rheology is not well
characterised with the used surfactant.

This work opens many perspectives. Other control parameters remain
to be studied, like polydispersity and rheological properties of
the surfactants. The effects of those parameters on the drag could
help to study their influence on foam rheology, and to explain the
scaling of the pressure gradient. Furthermore, the use of
asymmetric obstacles such as airfoil profiles will enlarge the
range of studied effects, allowing lift, torque or orientation
under flow. Pressure drop measurements, allowing to study
dissipation in foams \cite{Cantat2004}, are also currently
investigated by the authors. Now, a local analysis of the local
stresses, deformations \cite{Aubouy2003} and velocity fields is
required to provide a more detailed comprehension of the foam
rheology. The comparison between this local analysis and the
global properties of the foam, such as our drag measurements,
could provide a way to propose and test constitutive equations for
the mechanics of foams.

\begin{ack}
The authors would like to thank Franck Bernard, Kamal Gam and Julien Deffayet
for experimental help, the machine shop of Laboratoire de Spectrom\'etrie
Physique and Patrice Ballet for technical support, and Simon Cox, Wiebke
Drenckhan, Isabelle Cantat and Renaud Delannay for enlightening discussions.
\end{ack}

\end{document}